# Observation of supercavity modes in subwavelength dielectric resonators

Mikhail Odit[1,2,†], Kirill Koshelev[1,3,†], Sergey Gladyshev[1],

Konstantin Ladutenko[1], Yuri Kivshar[1,3,*] and Andrey Bogdanov[1,+]

[1]Department of Physics and Engineering, ITMO University, St Petersburg 197101, Russia
[2] Electrotechnical University LETI, St Petersburg 197376, Russia
[3]Nonlinear Physics Center, Research School of Physics, Australian National University,
Canberra ACT 2601, Australia

**Electromagnetic response of dielectric resonators with high refractive index is governed by optically induced electric and magnetic Mie resonances facilitating confinement of light with the amplitude enhancement[1,2,3]. However, strong subwavelength trapping of light was associated traditionally only with plasmonic or epsilon-near-zero structures which however suffer from losses. Recently, an alternative localization mechanism was proposed to trap light in individual subwavelength optical resonators with a high quality factor[4–6] in the regime of a supercavity mode[7]. Here, we present the experimental observation of the supercavity modes in subwavelength ceramic resonators in the radiofrequency range. We demonstrate experimentally that the regime of supercavity mode can be achieved via precise tuning of the resonator's dimensions resulting in a huge growth of the quality factor reaching the experimental values up to $1.25\times10^4$, being limited only by material losses in dielectrics. We reveal that the supercavity modes can be excited efficiently by both near- and far-fields by means of dipole sources and plane waves, respectively. In both the cases, the supercavity mode manifests itself clearly via characteristic peculiarities of the Fano resonance and radiation patterns. Our work paves the way for future compact practical devices in photonics and radiophysics.**

Enhancement of electromagnetic field is one of the key problem of optics, radiophysics, photonics and related branches of science. Therefore, the light trapping structures with high quality factor (Q factor) are vital for photonics industry and technologies. Optical resonators and microcavities are based on materials or systems that forbid outgoing waves by different means. The common principle of operation for traditional resonators and microcavities[8] is due to reflection from the resonator's boundaries under the condition of constructive interference of the trapped waves with themselves. For such structures, high Q factors usually demand large resonator size. The Q factor can be increased by engineering the environment as for cavities in photonic bandgap structures[9] or by exploiting the total internal reflection at glancing angles of incidence in whispering-gallery-mode resonators[10–12] For compact geometries, the lowest electric and magnetic Mie resonances dominate the optical response explaining why the value of the Q factor drops rapidly with a decrease of size. The creation of compact resonators with high Q factor is still a challenge.

† equal contribution
* ysk@internode.on.net
+ a.bogdanov@metalab.ifmo.ru

Recently, a new prospective mechanism was proposed for achieving high values of Q factor in high-refractive-index resonators of subwavelength scale[4,5]. The mechanism allows to achieve high-quality (high-Q) modes, or *supercavity* modes[7], in a single dielectric resonator via destructive interference of a pair of leaky modes with similar far-field profiles. The interference is enabled by continuous tuning of resonator's aspect ratio which induces strong coupling of the leaky modes when their frequencies come close to degeneracy[6]. The physics of supercavity modes was associated with bound states in the continuum (BICs) – exotic nonradiative states proposed in quantum mechanics almost a century ago[13] and re-discovered in photonics only in 2008[13]. Until now, optical BICs have been predicted and observed essentially in dielectric structures, such as photonic crystals[15,16], waveguide arrays[17] and metasurfaces[18], and used for various applications including lasing[19,20], biosensing[21] and nonlinear frequency conversion[4].

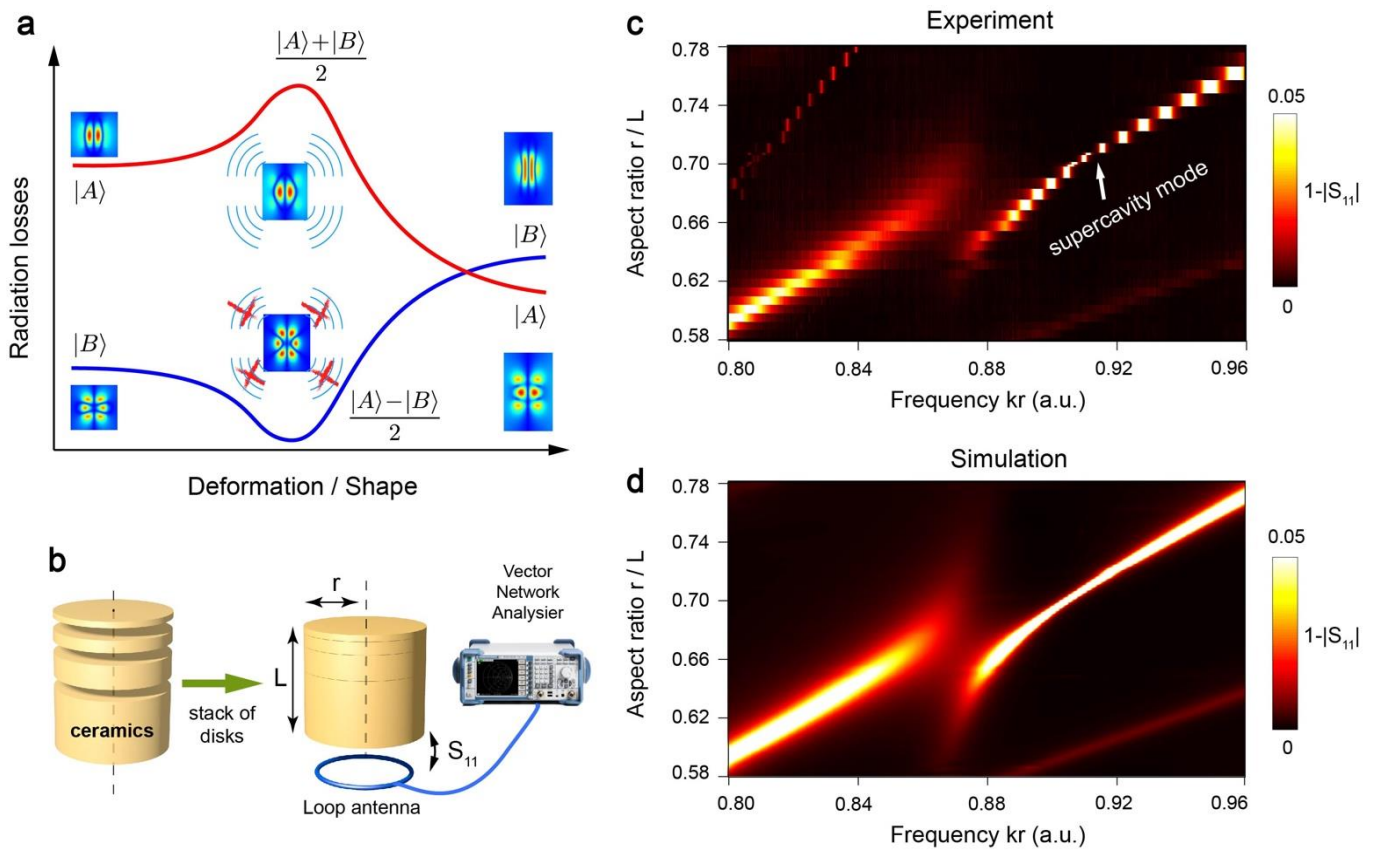

**Figure 1. Supercavity modes in subwavelength resonators and their experimental observation.** (a) Evolution of radiation losses for interacting optical modes $|A\rangle$ and $|B\rangle$ of a disk resonator in the strong coupling regime. The mode marked with blue line exhibits strong suppression of radiation losses under changing of the resonator's size and formation of the high-Q supercavity mode. (b) Experimental setup for the scattering measurements at radiofrequency range. The resonator consists of several ceramic disks of different height stacked in a single disk. The disk has the same radius $r$ =15.7 mm, permittivity $\varepsilon$ = 44.8 and loss tangent of the ceramics about tan $\delta$ = $10^{-4}$. The resonator is excited by a loop antenna connected to a vector network analyser. (c-d) Map of the measured and calculated scattering coefficient 1-|S$_{11}$| vs frequency *kr* and aspect ratio *r/L*, which demonstrates a clear avoid crossing between two resonances accompanied by supercavity mode appearance.

The interaction between two resonances with finite lifetimes results in a mixing of the resonant states and appearance of an avoided resonance crossing of their energies in the parameter space - the hallmark of the strong coupling regime. For closed systems with interaction, the losses of the dressed states are always between the losses of the bare states – eigenstates without interaction[22]. For open systems the situation is essentially different because the modes with a common radiative channel can couple externally via constructive and destructive interference. The lifetime of each interacting mode changes dramatically in this case. Thus, the constructive interference gives rise to super-radiant modes with enhanced radiation and the destructive interference can result in appearance of the mode with completely damped radiative losses – bound state in the continuum (BIC) – that was shown by Friedrich and Wintgen for quantum mechanical systems[23]. For a true BIC, the radiative losses through all the channels are completely damped making the mode dark and optically inactive. In optics BICs can exist only in unbounded structures or in structures with epsilon-near-zero materials, while for finite samples the complete loss cancelation is impossible[13]. Nevertheless, we can substantially increase the Q factor of the resonator by cancellation of radiation through the main radiative channels. This regime can be associated with supercavity mode also known as quasi-BIC. Recently, it was shown that the change of the resonator's size can play a role of the interaction between the modes and the radiation losses can be suppressed dramatically via continuous deformation indicating formation a high-Q supercavity mode[4,5,24] (see Fig. 1a). Very recently, the supercavity modes were proposed to realize a compact semiconductor nanolaser[25].

To observe a supercavity mode we manufacture a ceramic cylindrical resonator with the permittivity $\varepsilon = 44.8$ and loss tangent of the ceramics about $10^{-4}$. The resonator consists of a stack of the disks of different heights and the same radii *r* =15.7 mm (Fig. 1b). This design makes it possible to gradually change the height of the resonator and consequently its aspect ratio. The thickness of the thinnest disk defining the height change step is 0.2 mm and the thickest one is 15 mm. The resonator was placed on a holder made of dielectric foam with the permittivity approximately $\varepsilon = 1.1$ and low material losses at GHz frequencies. The cylindrical resonator hosts radial modes (Mie modes) and axial modes (Fabry-Perot modes), which electromagnetic fields can be characterized with an azimuthal degree of freedom (orbital angular momentum). We focus only on the azimuthal symmetric modes with zero angular momentum since they possess Q factors higher that for less symmetric modes. To excite the azimuthal symmetric modes selectively we use a loop antenna placed beneath the resonator concentrically with its axis. To study the scattering properties, we use the vector network analyzer (VNA) and measure the complex reflectivity via S$_{11}$ coefficient of the scattering matrix. The measured dependence of 1-|S$_{11}$| on the frequency *kr* and disk aspect ratio *r/L* demonstrates the avoided resonance crossing behavior with a characteristic linewidth narrowing for the high-frequency mode (see

Fig. 1c). The numerical simulations with identical excitation conditions show perfect coincidence with the experiment (see Fig. 1d).

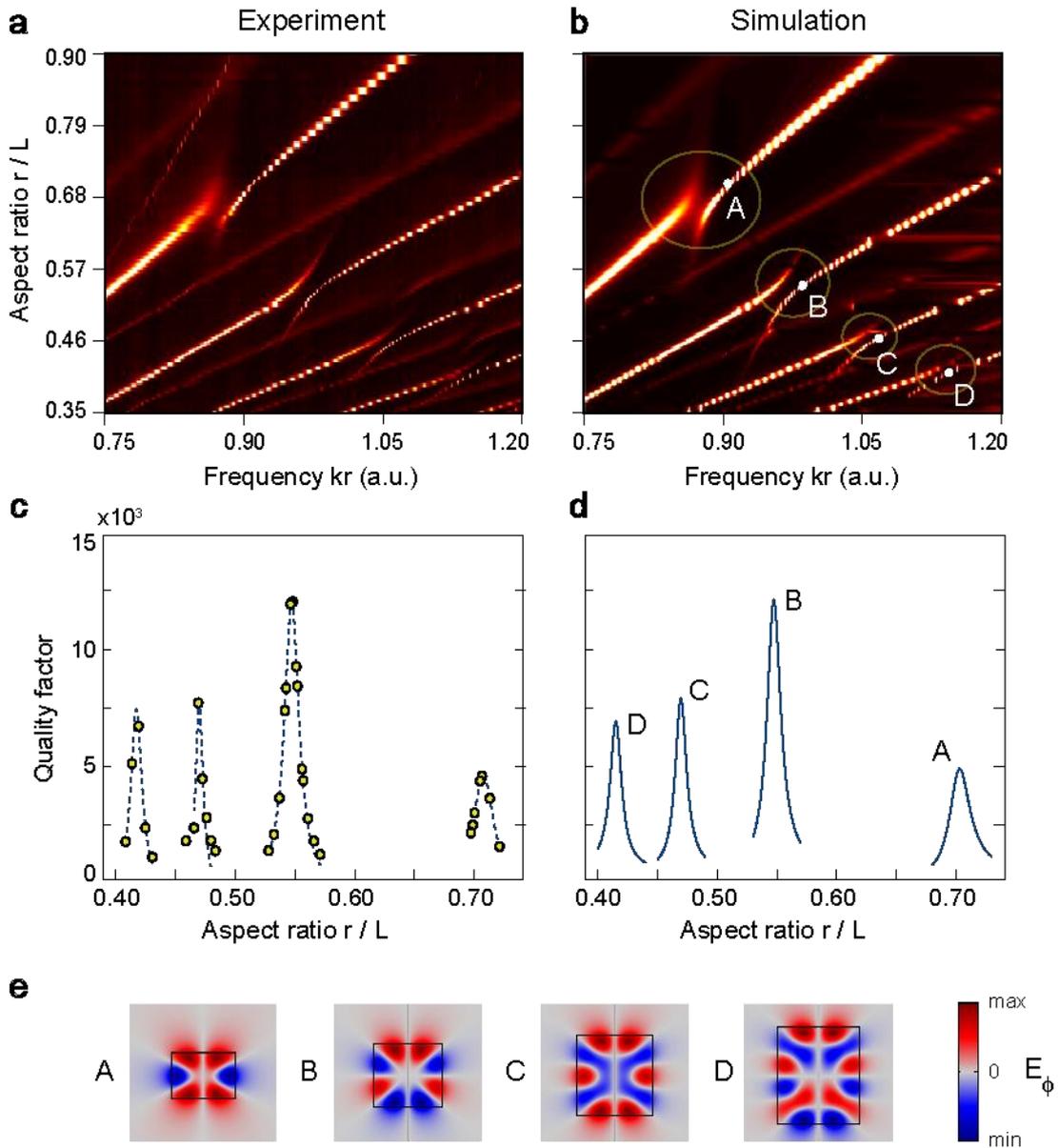

**Figure 2. Measurements of the Q factors for supercavity modes.** (a, b) Measured and calculated map of the scattering coefficient 1-|S$_{11}$| vs frequency kr and aspect ratio r/L. The supercavity modes labelled as A-D correspond to aspect ratios r/L = 0.71, 0.55, 0.7, 0.42. Measured (c) and calculated (d) dependence of Q factor on aspect ratio *r/L* for supercavity modes in the vicinity of the avoided resonance crossings. (e) Calculated near-field patterns of electric field for the supercavity modes (points A-C).

With the same setup we perform measurements of the reflectivity S$_{11}$ for wider range of cylinder aspect ratio and observe four supercavity modes (see Fig. 2a). Using the fitting procedures described in supplementary part, we extract the experimental unloaded Q factors of the modes and study their dependence on the disk aspect ratio in the vicinity of each avoid crossing (see Fig. 2b). The supercavity mode labeled as B possesses the highest Q factor of 12500 which matches perfectly with the numerical simulations (see Figs. 2c and 2d). For the numerical simulations in Fig.2d we apply an expansion of Maxwell's equations into the basis of resonant states (quasi-normal modes)[26,27]. This value of the Q factor is limited mainly by the material loss in the ceramics (loss tangent about $10^{-4}$). The direct finite element simulations show that the pure radiative Q factor reaches a value around 180 000. To get this value with direct measurements we need to use a resonator made of a material with loss tangent less than $6 \times 10^{-6}$ which is not possible with ceramic materials. The calculated near-field electric profiles for the supercavity modes show highly symmetric distributions with the increase of axial and radial order for the high-frequency modes.

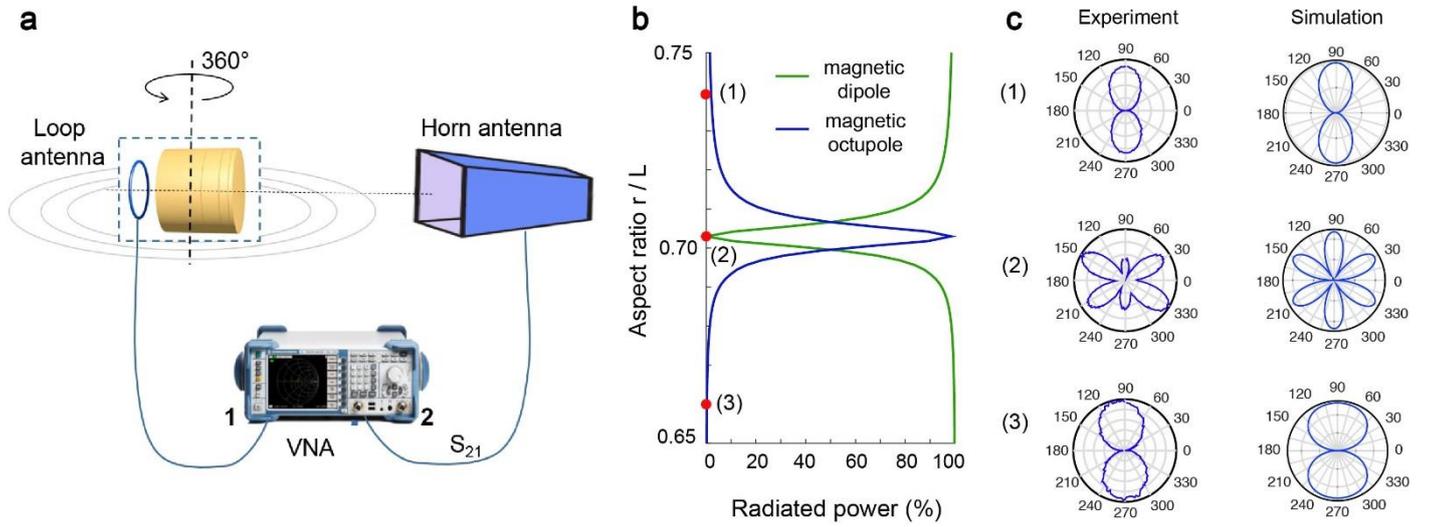

**Figure 3. Far-field measurements.** (a) Experimental setup for the far field radiation pattern measurements. Symmetrical modes of the resonator are excited by the loop antenna connected to the port 1 of VNA. Resonator and loop antenna rotate 360 degree around their axis. The scattered field collected by the horn antenna connected to the port 2 of VNA. (b) Contribution of the magnetic dipole and magnetic octupole to the radiated power of "A" supercavity mode (see Fig. 2) depending on aspect ratio r/L. (c) Normalized measured and calculated far field radiation patterns for three values of aspect ratio r/L = 0.66, 0.7025, 0.74. The amplitude of the scattered field increases drastically in the *supercavity*-regime (the frequency is 2.768 GHz).

In order to gain deeper insight into the physics of supercavity modes in individual high-index resonators, we illustrate cancelation of its radiative losses through the dominant scattering channel in term of multipoles. As was shown theoretically[6], in the supercavity regime the radiation through the dominant multipole channel becomes negligible. We perform the experimental study to demonstrate cancelation of supercavity mode radiation losses through the dominant channel in far-field. In the experiment we excited the cylindrical resonator with the loop antenna placed concentrically to the axis of the cylinder. The resonator together with the loop antenna rotated around transvers axis as shown in Fig. 3a. The scattered field in the orbital plane was measured by the distantly positioned horn antenna. We observe that the appearance of a supercavity mode "A" (Fig. 2b) is accompanied by a drastic change in the far-field radiation pattern from the magnetic dipolar to the magnetic octupolar one (see Fig. 3c) that agrees with theoretical predictions in Ref. 4.

It is well-known that the scattering spectra of a finite size obstacle can be described by a cascade of Fano resonances[28]. Fano parameters are used widely to approximate the experimental scattering spectra and they are considered as independent fitting parameters. It was predicted that in compact structures, the supercavity modes are fundamentally linked to optical Fano resonances[6]. In particular, it was shown theoretically that the Fano asymmetry parameter diverges in the vicinity of the supercavity regime, when the Q factor as a function of the aspect ratio reaches the maximum. This prediction has a very powerful practical meaning – a Fano resonance in scattering spectrum as function an external parameter reaches the maximal Q factor when it turns into the symmetric Lorentzian shape. This prediction was not verified experimentally.

To confirm it we investigate experimentally the scattering properties of the cylindrical resonator in a radiofrequency range. The resonator was positioned between two horn antennas connected to the ports of vector network analyzer. The first antenna illuminated quasi plane wave in the direction of cylinder. The field is polarized orthogonally to the axis of the cylinder. The second horn antenna collected the field scattered in the front direction. The resulted amplitude of the transmission coefficient was used to calculate the total extinction cross section by means of the optical theorem[29]. The measured and calculated maps of the extinction cross-section vs aspect ratio $r/L$ and dimensionless frequency $kr$ are plotted inn Figs. 4e and 4f, respectively. The measured and numerically calculated extinction cross-sections (see Figs. 4b and 4c) were used for the extraction of the Fano asymmetry parameter (Fig. 4d). One can clearly see that the Fano asymmetry parameter diverges in the vicinity of the supercavity modes serving as clear indicator of its manifestation along with growing the Q factor and changing the far-field radiation pattern.

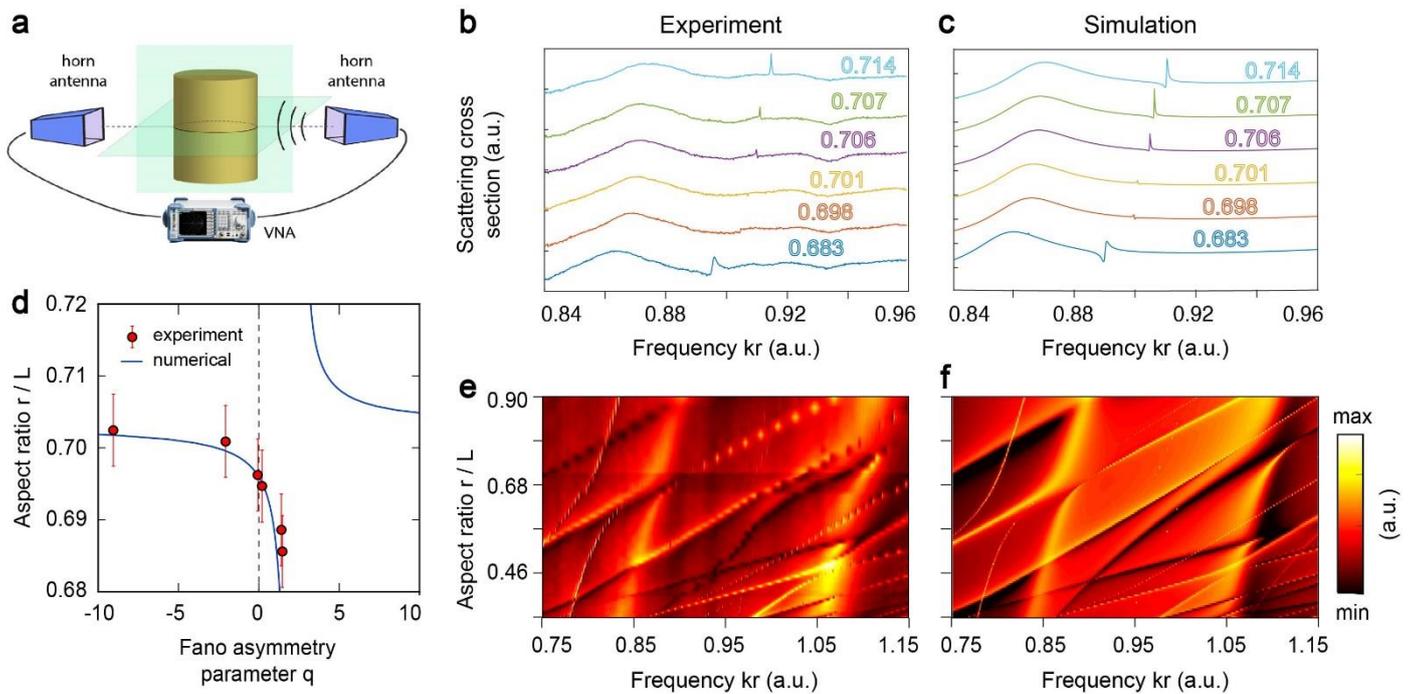

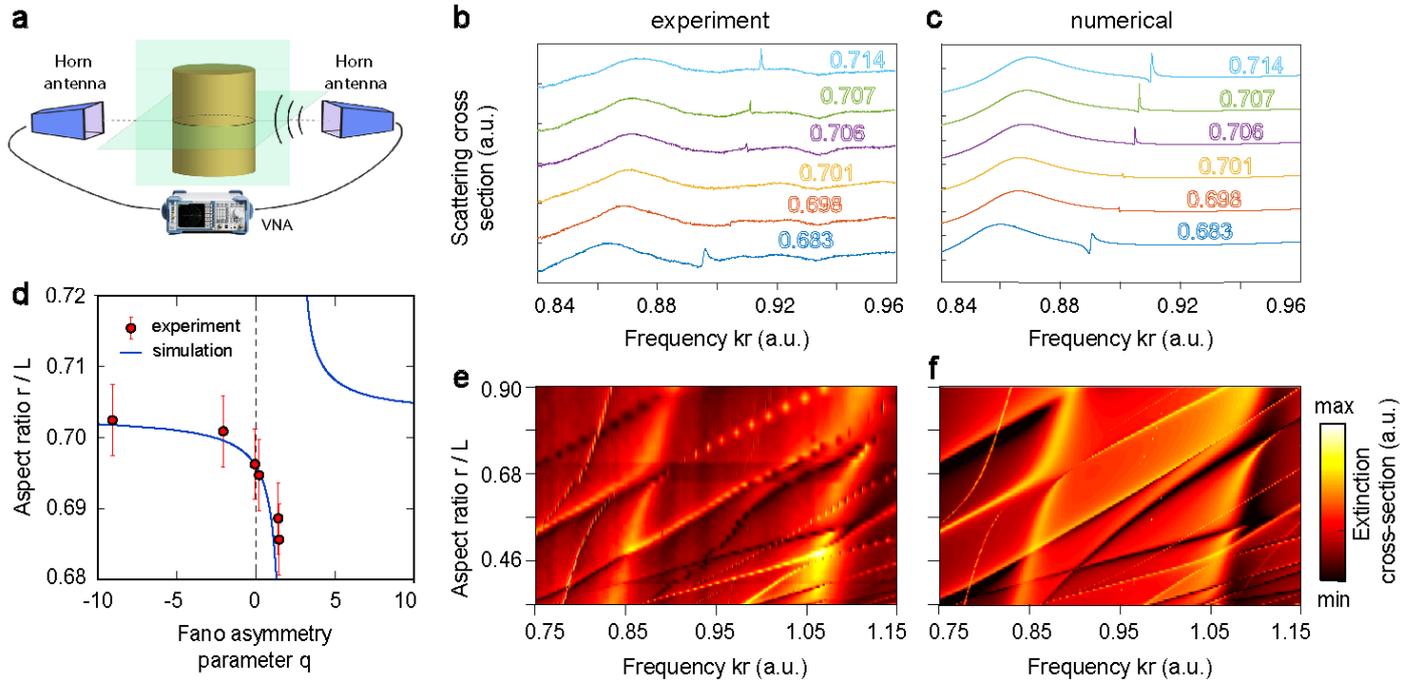

**Figure 4. Characterization of the Fano asymmetry parameter.** (a) Experimental setup for the forward scattering measurement. (b, c) Measured and calculated spectra of the normalized total scattering cross-section as a function of resonator aspect ratio in the region of the avoided resonance crossing "A". (d) Extracted and simulated Fano asymmetry parameter depending on r/L. (e, f) Experimental and numerical dependencies of the total scattering cross-section on the aspect ratio of the cylinder and frequency for TE-polarized incident wave. The cross section is normalized to the projected cross-section S=2rL.

In summary, we have demonstrated experimentally the excitation of supercavity modes in high-index dielectric resonators in the subwavelength regime in the radiofrequency range. Such high-Q resonant electromagnetic states of individual open dielectric resonators appear due to interference of two dissimilar leaky modes excited simultaneously, and they are manifested in peculiarities in the scattering cross-section spectra. We have observed in experiment that the supercavity mode are recognized clearly through a dramatic change in the far-field radiation pattern of the electromagnetic fields.

**METHODS SUMMARY**

**Sample fabrication.** The dielectric cylindrical resonator was composed of the ceramic disks of different heights and same diameters. Each disk has been manufactured by the sintering of the ceramic powder of the calcium titanate-lanthanum aluminate ($LaAlO_3$–$CaTiO_3$) solid solution in a cylindrical form. Each disk has been than polished in order to get the required height (from 0.25 mm to 15mm).

The dielectric foam holder for the resonator has been fabricated by the CNC (computer numerical control) machine drilling of the Penoplex© foam material.

**Numerical calculations.** Reflection coefficients in Figs. 1d and 2b have been calculated by the frequency domain solver of the CST Microwave Studio. The dielectric resonator has been excited by the loop antenna placed 7 mm below its base. The loop is excited by the discrete port. The boundary conditions are open (add space) from all sides. The near-field patterns for the supercavity modes of the dielectric resonator were calculated using COMSOL software. The scattering properties (Figs. 4c and 4f) are calculated with the time domain solver of the CST Microwave Studio. The dielectric resonator has been excited by the plane wave polarized orthogonally to the resonance cylinder geometrical axis. The total radar cross section has been calculated with a broadband far field monitor and normalized to the geometrical cross section of the resonator.

**Microwave measurements.** The reflection scattering of the dielectric resonator was obtained by measuring the reflection coefficient of the loop antenna placed below the basis of the cylindrical resonator. The vector network analyser Rohde & Schwartz ZVB-40 is used to measure $S_{11}$ reflection coefficient. The far field radiation pattern of the cylindrical dielectric resonator excited by the loop antenna measured by the linearly polarized wideband horn antenna (operational range 1 - 18 GHz) connected to the ports of an Agilent E8362C Vector Network Analyzer. Measurements performed in an anechoic chamber with subsequent time gating in order to perform noise reduction. The background signal was subtracted by means of free space measurement. The scattering cross section was obtained by measuring the transmission between two horn antennas placed in line with the cylindrical resonator. The total scattering cross section was calculated by means of optical theorem.

**Fitting procedures.** We use custom scripts in Python programming language to extract resonances from experimental data. Fit of the Fano parameters was done using Levenberg-Marquardt algorithm from SciPy software library. Extraction of unloaded Q-factor from complex $S_{11}$ was done using QZERO software[30] kindly provided by Prof. D. Kajfez.

**Acknowledgements**

The authors thank M. Limonov and M. Rybin for useful discussions and collaboration. K.K. and Y.K. acknowledge a financial support from the Australian Research Council and Strategic Fund of the Australian National University. K.K. and A.B. acknowledge a support from the Foundation for the Advancement of Theoretical Physics and Mathematics "BASIS". M. O. acknowledges the support from the RFBR (grant 18-37-00486).

**Author contributions**

A.B. and Y.K. conceived this research project. K.L., S.G. and K.K. performed numerical calculations and theoretical studies. M.O. fabricated the samples and collected experimental data. M.O. and K.K. wrote the manuscript based on the input from all the authors.

**Competing financial interests**

The authors declare no competing financial interests.

Correspondence and requests for materials should be addressed to Andrey Bogdanov or Yuri Kivshar.